\documentclass[]{spie}  

\usepackage{amsmath,amsfonts,amssymb}
\usepackage{graphicx}
\usepackage[colorlinks=true, allcolors=blue]{hyperref}

\usepackage{bm}
\usepackage[ruled]{algorithm2e}

\newcommand{\cd}{c^{\dagger}}
\newcommand{\tR}{t_{1R}}
\newcommand{\tl}{t_{1L}}
\newcommand{\ee}{{\rm{e}}}
\newcommand{\ii}{{\rm{i}}}

\newcommand{\jr}{J_R^{\dagger}}

\title{Four-fold non-Hermitian phase transitions and non-reciprocal coupled resonator optical waveguides}

\author[a]{Xintong Zhang, Xiaoxiao Song, Shubo Zhang, Tengfei Zhang, Mengqi Lv, Yiyun Zou, Jing Li}
\affil[a]{Department of Optical Science and Engineering, Shanghai Ultra-Precision Optical Manufacturing Engineering Center, Fudan University, Shanghai 200438, China}

\authorinfo{Jing Li: E-mail: lijing@fudan.edu.cn}

\begin{document} 
\maketitle

\begin{abstract}
Non-Hermitian systems can exhibit extraordinary sensitivity to boundary conditions. Given that topological boundary modes and non-Hermitian skin effects can either coexist or individually appear in non-Hermitian systems, it is of great value to present a comprehensive non-Hermitian phase diagram, for further flexible control in realistic non-Hermitian systems. Here, we reveals four-fold non-Hermitian phase transitions at a mathematically level, where phase I exhibits only topological boundary modes, phase II displays both topological boundary modes and skin modes, phase III exhibits only skin modes, and phase IV cannot manifest any boundary modes. By deriving non-Hermitian winding numbers, the existence or non-existence condition of topological boundary modes are analytically expressed, consistent with the numerical results obtained through the iterative Green’s function method. Combining with the study on non-Hermitian skin effects, we rigorously establish the four-fold phase diagram. We also design an array of coupled resonator optical waveguides. The introduction of non-Hermiticity in the photonic structure induces a phenomenon similar to band inversion in topological insulators, indicating the presence of topological boundary modes in the photonic bands.

\end{abstract}


\section{INTRODUCTION}
\label{sec:intro}  

Band topology in non-Hermitian physics has attracted in recent years\cite{Mostafazadeh2002,Bender2007,El-Ganainy2018,Gong2018,Ghatak2019,Ashida2020,Bergholtz2021}. As is well-known for Hermitian systems, a nontrivial Bloch band invariant implies the emergence of topological boundary modes under open boundary condition (OBC)\cite{Bernevig2006,Schnyder2008}, which can date back to the discovery of integer quantum Hall effect\cite{Klitzing1980,Thouless1982,Hatsugai1993}. The interplay between non-Hermitian physics and topological physics extends the richness of the phase transitions. On one hand, traditional bulk-boundary correspondence appears to be violated in non-Hermitian regimes, where the topological phase transition points are not consistent with the bulk gap-closing points\cite{Lee2016,Yao2018,Kunst2018,Lee2019}. On the other hand, non-Hermitian skin effect (NHSE) phenomenon\cite{Yao2018a,Yao2018,Yokomizo2019,Kawabata2020,Guo2021b} is observed, where all the eigenstates under OBC are exponentially localized at the boundary of the lattice. Inspired by these phenomena, non-Hermitian topology provides higher reconfigurability in non-Hermitian classical systems or quantum systems, such as photonics\cite{Zhu2020,Zhong2021,Chen2021,Wang2021,Parto2021}, electrical circuits\cite{Yoshida2020,Hofmann2020,Xu2021,Zhang2021,Zou2021}, acoustics\cite{Acoustics2021}, magnetics\cite{Flebus2020,Deng2022}, and quantum-walk dynamics\cite{Xiao2020}, which are beyond any known phenomenon from their Hermitian counterparts. 

Non-Hermitian topological photonics has attracted widespread attention in recent years\cite{Ozawa2019,Nasari2023}. The inherent gain and loss characteristics of photonic systems provide a foundation for validating non-Hermitian theory. Relevant theoretical and experimental studies indicate that, artificial optical lattices, composed of coupled resonator optical waveguides (CROW)\cite{Poon2004,Dar2006,Zhu2020}, provide a flexible platform to study non-Hermitian theory, such as topological phase transitions in PT-symmetric CROW system\cite{Weimann2017,Ao2020}, anomalous Floquet NHSE in a ring resonator lattice\cite{Gao2022}, V-shaped NHSE\cite{Xin2023}, and higher-order NHSE\cite{Chen2021}. Meanwhile, the non-Hermitian topology band theory has introduced innovative mechanisms for manipulating optical fields in the field of photonics, such as single-mode lasing induced by nonlinearity and NHSE\cite{Zhu2022}, and near-field beam steering in a Hatano–Nelson laser array\cite{Liu2022}.

Given that topological boundary modes and non-Hermitian skin effects can either coexist or individually appear, it is of great value to present a comprehensive non-Hermitian phase diagram, for further flexible control in realistic non-Hermitian systems. In this paper, we explore various non-Hermitian phases at a mathematically level, and investigate topological phase transition in non-Hermitian coupled resonator optical waveguides. In Section 2, we introduce a non-reciprocal Su-Schrieffer-Heeger (SSH) chain model featuring long-range couplings. This model reveals four-fold non-Hermitian phases: Phase I exhibits only topological boundary modes, phase II displays both topological boundary modes and skin modes, phase III exhibits only skin modes, and phase IV cannot manifest any boundary modes. By deriving non-Hermitian winding numbers, the existence or non-existence condition of topological boundary modes are analytically expressed, which corresponds to the numerical outcomes obtained through the iterative Green’s function method. Combining with the study on non-Hermitian skin effects, we rigorously establish the four-fold phase diagram. In Section 3, we design an array of coupled resonator optical waveguides, and explore non-Hermitian phases within this photonic structure. Achieving non-reciprocity is facilitated by link rings incorporating both gain and loss materials. Results obtained from the transfer matrix method reveal that, the introduction of non-Hermiticity in our photonic structure induces a phenomenon similar to band inversion in topological insulators, indicating the presence of topological boundary modes in the photonic bands. This work theoretically has significant guiding implications for controlling the non-Hermitian phase transition characteristics in real physical systems, introducing a new control mechanism to the field of photonics.   


\section{Non-Hermitian Phase Transitions}
\subsection{Numerical Results}

\begin{figure}[htb]
\centering\includegraphics[width=0.4\linewidth]{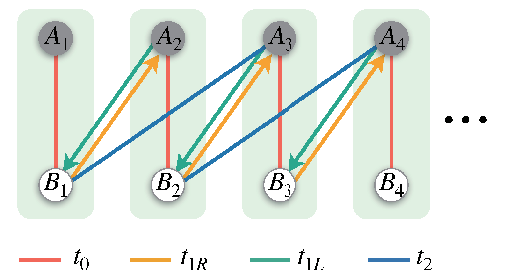}\caption{\label{fig1_lattice}{\bf Schematic view of the one-dimensional non-Hermitian lattice model with long-range couplings}. Within each unit cell, there are two sites denoted as $A$ and $B$. The hopping parameters $t_{1R,1L}$ represent the non-reciprocal nearest inter-cell interaction, while $t_0$ and $t_2$ represent the reciprocal intra-cell and next-nearest inter-cell interaction.}
\end{figure}

Here, we begin with the tight-binding model of our one-dimensional non-Hermitian chain model featuring long-range couplings. As shown in Fig.\ref{fig1_lattice}, each unit cell consists of two inequivalent sites, $A$ and $B$. The hopping term inside each cell is represented by $t_0$, while the hopping terms between the nearest cells are represented by $t_{1R},t_{1L}$, and the hopping term between the next-nearest cells is represented by $t_2$. For simplicity and without generality, all hopping parameters are assumed to be real numbers. Consequently, the non-Hermiticity of our chain model depends on the non-reciprocity of the inter-cell interaction, i.e., $t_{1R}\neq t_{1L}$.

According to the schematic view illustrated in Fig.\ref{fig1_lattice}, the Hamiltonian of our chain model in second quantization can expressed as,
\begin{equation}
\begin{split}
\hat{H}= &\sum_{n=1}^N [\varepsilon_0(\cd_{n,A}c_{n,A}+\cd_{n,B}c_{n,B})+
(t_0\cd_{n,A}c_{n,B}+h.c.)]\\
&+\sum_{n=1}^{N-1}\tR\cd_{n+1,A}c_{n,B}+\tl\cd_{n,B}c_{n+1,A}\\
&+\sum_{n=1}^{N-2}(t_2\cd_{n+2,A}c_{n,B}+ h.c.).
\end{split}
\label{real-space Hamiltonian}
\end{equation}
Here, $N$ is the total number of the unit cells, and $\varepsilon_0$ denotes the on-site energy of site $A$ and $B$. Open boundary condition is imposed in Eq.(\ref{real-space Hamiltonian}) to study boundary behaviors, where the hopping terms between $(1,N),(1,N-1),(2,N)$ cells are forbidden. 

\begin{figure}[htb]
\centering\includegraphics[width=1\linewidth]{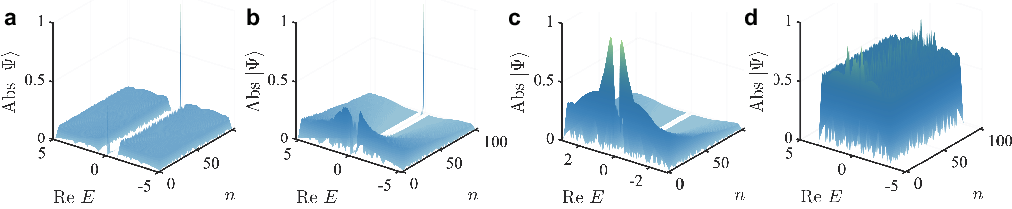}
\caption{\label{fig2_psi}{\bf Numerical results on the absolute value of eigenstates $|\Psi\rangle$ under open boundary conditions}. $|\Psi\rangle$ is as a function of the real part of eigenvalues $E$ and cell position $n$, where the on-site energy is set as $\varepsilon_0 = 0$ and $n=1,2,...,100$. (a) Phase I with hopping parameters $t_0 = 1$, $\tR =3.5 $, $\tl =2.5 $, $t_2 = 1$. (b) Phase II with hopping parameters $t_0 = 1$, $\tR = 3.5$, $\tl = 2.5$, $t_2 = 1.3$. (c) Phase III with hopping parameters $t_0 = 1$, $\tR = 1.2$, $\tl = 1.6$, $t_2 = 0.6$. (d) Phase IV with hopping parameters $t_0 = 1$, $\tR =1.2 $, $\tl = 1.6$, $t_2 = 1$.}
\label{fig_numerical}
\end{figure}

The numerical results of eigenvalues and eigenstates under open boundary conditions are obtained through the diagonalization of the Hamiltonian described in Eq.(\ref{real-space Hamiltonian}). As shown in Fig.\ref{fig2_psi}, a total of $N=100$ lattice cells is considered. The $x$-axis represents the spatial coordinate of the $n$th lattice cell, the $y$-axis represents the real part of the eigenvalues $E$, and the $z$-axis represents the absolute values of all the $2N=200$ eigenstates $|\Psi\rangle$. By adjusting the values of the hopping parameters $t_0,t_{1R},t_{1L},t_2$, a total of four distinct non-Hermitian phases can be observed. Among them, Fig.\ref{fig2_psi}(a) illustrates non-Hermitian phase I, characterized by the presence of topological boundary modes at $E = \varepsilon_0$, without non-Hermitian skin effects. As shown in Fig.\ref{fig2_psi}(b), non-Hermitian phase II exhibits both topological boundary modes and non-Hermitian skin effects. The skin modes are locally confined to one side of the lattice model at $n=1$. In Fig.\ref{fig2_psi}(c), non-Hermitian phase III displays only non-Hermitian skin effects without the presence of topological boundary modes. Finally, Fig.\ref{fig2_psi}(d) illustrates non-Hermitian phase IV, where none of the eigenstates exhibit any form of boundary modes, and all eigenstates are uniformly distributed.

\begin{figure}[htp]
\centering\includegraphics[width=0.6\linewidth]{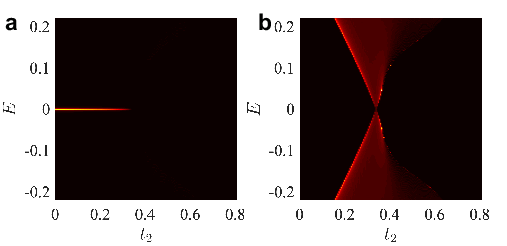}\caption{\label{fig3_green} {\bf Spectral functions calculated from the iterative Green's function method}. (a) Surface spectral function $A_{s,s'}(E)$. (b) Bulk spectral function $A_b(E)$. The hopping parameters are set as $t_0 = 1$, $\tR = 1.2$, $\tl = 1.6$ and $t_2 \in [0,0.8]$.}
\end{figure}

The iterative Green's function method\cite{Green2018} is also utilized for the numerical investigation of topological phase transition under the thermodynamic limit that $N\to\infty$. This method imposes the condition that only nearest-neighbor cell interactions exist in the lattice model. Therefore, it is essential to treat the one-dimensional chain model as a lattice with a total of $N/2$ supercells, where each supercell encompasses the original two cells. The detailed procedure of Green's function calculation is provided in Appendix \ref{sec_green}. Finally, the surface spectral function and bulk spectral function are demonstrated in Fig.\ref{fig3_green}(a) and (b), respectively. In the parameter space $t_2$, it is evident that the topological phase transition point for our one-dimensional chain model occurs at approximately $t_2\approx 0.34$. Since the average number of iteration is 8, this is equivalent to calculating a one-dimensional chain lattice model with a total of $N=2^8=512$ cells. Theoretically, the topological phase transition point obtained in the calculation is expected to be closer to the topological phase transition point under thermodynamic limit conditions.

\subsection{Non-Hermitian Winding Numbers}
To rigorously explore non-Hermitian phase transitions and establish the non-Hermitian bulk-boundary correspondence at a mathematically level, we derive non-Hermitian winding numbers\cite{Lee2019} to predict the topological phase transition points. Here, we start from the Hamiltonian $H(k)$ in momentum space, expressed as,
\begin{equation}
\begin{aligned}
H(k) & =
\begin{pmatrix}
0 &h_a(k)\\
h_b(k) & 0
\end{pmatrix},\\
\end{aligned}
\label{Hk}
\end{equation}
where $h_a(k)  =t_0+\tR {\ee}^{-{\ii}k}+ t_2{\ee}^{-{\ii}2k}$ and $h_b(k) =t_0+\tl {\ee}^{{\ii}k}+ t_2{\ee}^{{\ii}2k}$. Therefore, our one-dimensional chain model is protected by the sub-lattice symmetry that $\sigma_zH(k)\sigma_z^{-1}=-H(k)$ if $\varepsilon_0 = 0$, satisfying the precondition of non-Hermitian winding numbers method. 

The crucial step in this method involves replacing the Bloch wave vector $k$ by a non-Bloch wave vector $z$, where $z = {\ee}^{{\ii}\tilde{k}}$ and $\tilde{k}\in\mathbb{C}$. 
The imaginary part of $\tilde{k}$ will contain all the information about the open boundary conditions. The general criterion for the appearance of topological boundary modes is expressed as,
\begin{equation}
\exists\ R\in\mathbb{R}_+, W_a(R)W_b(R)<0,
\label{criterion}
\end{equation}  
where $W_{a,b}(R)$ is the so-called non-Hermitian winding number, calculated as the difference between the number of zeros $Q_{a,b}$ and the number of poles $P_{a,b}$ encircled by the contour $|z|=R$ for $h_{a,b}(z)$, 
\begin{equation}
\begin{aligned}
W_{a}(R)&=\frac{1}{2\pi{\rm i}}\oint_{|z|=R}d[{\rm ln}h_{a}(z)]=Q_a-P_a,\\
W_{b}(R)&=\frac{1}{2\pi{\rm i}}\oint_{|z|=R}d[{\rm ln}h_{b}(z)]=Q_b-P_b.
\end{aligned}
\label{winding}
\end{equation}  
Assuming that hopping parameters are all positive real numbers, the criterion for the appearance of topological boundary modes becomes,
\begin{equation}
\begin{aligned}
\max\{|z_a^{(1)}|,|z_a^{(2)}|\}>\min\{|z_b^{(1)}|,|z_b^{(2)}|\},
\end{aligned}
\end{equation}  
where $z_{a,b}^{(j)}$ are the roots of $h_{a,b}(z)=0$, solved as
\begin{equation}
\begin{aligned}
z_a^{(1,2)} = \frac{-\tR \pm \sqrt{\tR ^2 - 4t_0t_2}}{2t_0},\\
z_b^{(1,2)} = \frac{-\tl\pm \sqrt{\tl ^2 - 4t_0t_2}}{2t_2}.
\end{aligned}
\end{equation}
Therefore, the topological boundary modes appear if and only if
\begin{equation}
|\frac{-\tR - \sqrt{\tR ^2 - 4t_0t_2}}{2t_0}|>|\frac{-\tl+ \sqrt{\tl ^2 - 4t_0t_2}}{2t_2}|.
\end{equation}

\begin{figure}[htb]
\centering\includegraphics[width=0.6\linewidth]{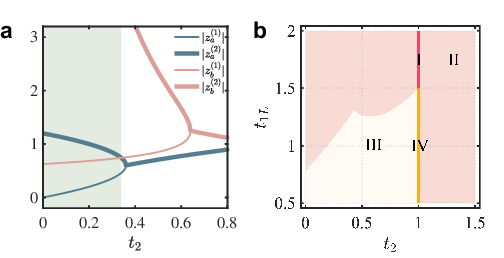}
\caption{\label{fig4_winding}(a) {\bf Non-Hermitian winding numbers}. The green lines represent the roots of $h_a(z)=0$, that is $z_a^{(1,2)} = ({-\tR \pm \sqrt{\tR ^2 - 4t_0t_2}})/{2t_0}$, while the red lines represent the roots of $h_b(z)=0$, that is $z_b^{(1,2)} = ({-\tl\pm \sqrt{\tl ^2 - 4t_0t_2}})/{2t_2}$. The light green region indicates the emergence of topological boundary modes where $|z_a^{(2)}|>|z_b^{(1)}|$. The hopping parameters are set as  $t_0 = 1$, $\tR = 1.2$, $\tl = 1.6$. (b) {\bf Non-Hermitian four-fold phase diagram}. Phase I exhibits only topological boundary modes, phase II displays both topological boundary modes and skin modes, phase III exhibits only skin modes, and phase IV cannot manifest any boundary modes. The hopping parameters are set as $t_0=1$ and $t_{1R}-t_{1L}=0.5$.}
\end{figure}

As shown in Fig.\ref{fig4_winding}(a), in the parameter space $t_2$, the topological phase transition point is rigorously derived as $t_2=0.3398$, which is consistent with the numerical result calculated from the iterative Green's function method. Furthermore, combining this with the condition of the non-existence of non-Hermitian skin effects, $t_0=t_2$, we establish a four-fold phase diagram for our chain model, as shown in Fig.\ref{fig4_winding}(b). In this non-Hermitian phase diagram, phase I, represented by the red line, indicates the presence of topological boundary modes and absence of non-Hermitian skin effects. Non-Hermitian phase II, denoted by the pink shaded area, signifies the coexistence of topological boundary modes and non-Hermitian skin effects. Non-Hermitian phase III, illustrated by the light yellow shaded area, exhibits only non-Hermitian skin effects without the presence of topological boundary modes. Non-Hermitian phase IV is represented by the yellow line segment, indicating the absence of any boundary modes. These analytical results are consistent with the numerical results shown in Fig.\ref{fig_numerical}. It is evident that, non-Hermitian phases I and IV maintain the traditional bulk-boundary correspondence observed in Hermitian systems, while non-Hermitian phases II and III possess non-Hermitian skin effects, a unique characteristic of to non-Hermitian systems.


\section{coupled resonator optical waveguides}
\begin{figure}[htb]
\centering\includegraphics[width=0.686 \linewidth]{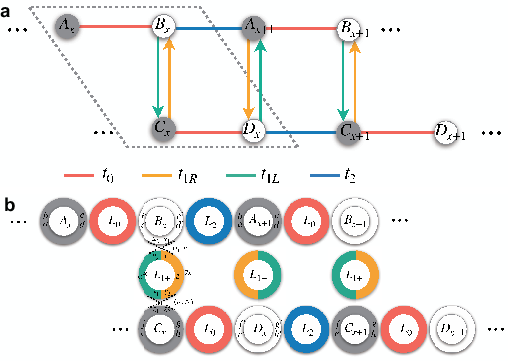}
\caption{\label{fig5_optical}(a) {\bf An equivalent form of our one-dimensional non-Hermitian chain model}. Each unit cell contains four sites $A,B,C,D$. (b) {\bf Schematic view of the one-dimensional coupled resonator optical waveguides}. The grey and white colored rings are site rings $A,C$ and $B,D$. The red and blue colored rings are link rings $L_0$ and $L_2$, without any gain or loss. In the link rings $L_{1\pm}$, the green semicircle represents the region with gain material, while the yellow semicircle represents the region with loss material.}
\end{figure}

Here, we have designed a coupled resonator optical waveguide and explored non-Hermitian phase transitions within this photonic structure. As illustrated in Fig \ref{fig5_optical}(a), our one-dimensional chain model is transformed into an equivalent tight-binding model without changing its topological characteristics. Each unit cell in this modified lattice model contains four sites $A,B,C,D$. Figure \ref{fig5_optical}(b) presents the schematic view of our coupled resonator optical waveguide, comprising  site rings and link rings. Among them, the site rings, colored in grey or white, correspond to sites $A,C$ or $B,D$, constructed from ideal dielectric materials without any gain or loss. The complex amplitude of the electric field for site rings $A,B,C,D$ is denoted as $\{a,b,c,d\},\{a',b',c',d'\},\{e,f,g,h\},\{e',f',g',h'\}$, respectively.  The red-colored link rings, denoted as $L_0$, correspond to the reciprocal hopping term $t_0$ within each cell, while the blue-colored link rings, denoted as $L_2$, correspond to the reciprocal hopping term $t_2$ between the nearest cell. Non-Hermitian terms require the use of link rings $L_{1\pm}$ with gain and loss materials, where the semicircular segments in green denote regions with gain materials, while the semicircular segments in yellow represent regions with loss materials.

Transfer-matrix method\cite{Poon2004,Dar2006} is employed to investigate the photonic band structure and transmission spectrum. This structure exhibits two mutually independent propagation modes within site rings, namely, clockwise and counterclockwise modes. In this context, we will focus on illustrating the clockwise mode. To begin, the transmission coefficient and coupling coefficient between link ring $L_{1\pm}$ and their coupled site rings are represented by $t_1\in \mathbb{R}$ and $r_1\in{\ii}\mathbb{R}$, where $t_1^2-r_1^2=1$. For the coupling between $L_{0}$/$L_2$ and their neighboring rings, the transmission and coupling coefficients are denoted by $t_0$/$t_2$ and $r_0$/$r_2$. The phase of the electric field will experience a change, denoted as $\phi$, after 1/4 propagation within the ring, where $\phi=\pi n L/2\lambda_0=\pi\omega/2$. Here, $n$ is the real part of the refractive index, $L$ is the length of ring, $\lambda_0$ is the designed wavelength, and $\omega$ is the normalized frequency. If the ring is constructed from materials with gain or loss, the amplitude of the electric field will change by ${\ee}^{\pm\kappa},\kappa>0$. Consequently, the effective transmission and coupling coefficients between two site rings can be derived as follows,
\begin{equation}
\begin{aligned}
T_{\alpha}=\frac{t_{\alpha}(1-\ee^{\ii 4 \phi})}{1-t_{\alpha}^2\ee^{\ii 4 \phi}} ,\
R_{\alpha}= \frac{r_{\alpha}^2\ee^{\ii 2 \phi}}{1-t_{\alpha}^2\ee^{\ii 4 \phi}} ,\ \alpha=0,1,2.
\end{aligned}
\end{equation} 
               
Owing to the Bloch theorem, the complex amplitudes of the electric fields satisfy the following relation,
\begin{equation}
\begin{aligned}
{\ee}^{{\ii} k_x}\left(\begin{matrix}
a_{x}'\\
b_{x}'\\
e_{x}\\
f_{x}
\end{matrix}\right)=Q_4Q_3Q_2Q_1
\left(\begin{matrix}
a_{x}'\\
b_{x}'\\
e_{x}\\
f_{x}
\end{matrix}\right).
\end{aligned}
\label{eigenQ}
\end{equation}
The transfer matrix $Q_1,Q_2,Q_3,Q_4$ can be written as,
\begin{equation}
\begin{aligned}
Q_1&=\left(\begin{matrix}
0					&\ee^{\ii 2\phi}	& 0 	& 0\\
\ee^{-\ii 2\phi}/{T_1}	&0			& 0	& -\ee^{-2\kappa}R_1/{T_1}\\
\ee^{2\kappa}R_1/{T_1}&0			& 0 	&  \ee^{\ii 2\phi}(T_1^2-R_1^2) /{T_1}\\
0					&0			& \ee^{-\ii 2\phi} & 0
\end{matrix}\right),
\end{aligned}
\end{equation} 
\begin{equation}
\begin{aligned}
Q_2&=\left(\begin{matrix}
-T_2/{R_2}			&1/{R_2}		& 0 	& 0\\
-(T_2^2-R_2^2) /{R_2}		&T_2/{R_2}	& 0 	& 0\\
0 						& 0 				& -T_0/{R_0}				&1/{R_0}\\
0 						& 0 				& -(T_0^2-R_0^2) /{R_0} 		&T_0/{R_0}
\end{matrix}\right),
\end{aligned}
\end{equation} 
\begin{equation}
\begin{aligned}
Q_3&=\left(\begin{matrix}
0					&\ee^{\ii 2\phi}	& 0 	& 0\\
\ee^{-\ii 2\phi}/{T_1}	&0			& 0	& -\ee^{2\kappa}R_1/{T_1}\\
\ee^{-2\kappa}R_1/{T_1}&0			& 0 	&  \ee^{\ii 2\phi}(T_1^2-R_1^2) /{T_1}\\
0					&0			& \ee^{-\ii 2\phi} & 0
\end{matrix}\right),
\end{aligned}
\end{equation} 
\begin{equation}
\begin{aligned}
Q_4&=\left(\begin{matrix}
-T_0/{R_0}			&1/{R_0}		& 0 	& 0\\
-(T_0^2-R_0^2) /{R_0}		&T_0/{R_0}	& 0 	& 0\\
0 						& 0 				& -T_2/{R_2}				&1/{R_2}\\
0 						& 0 				& -(T_2^2-R_2^2) /{R_0} 		&T_2/{R_2}
\end{matrix}\right).
\end{aligned}
\end{equation} 
Therefore,  under the given wave vector $k_x\in[-\pi,\pi]$, the values of the normalized frequency $\omega$ can be obtained by solving the equation $\det[Q_4Q_3Q_2Q_1 - \ee^{\ii k_x}\mathbb{I}]=0$, thereby determining the band structure.

\begin{figure}[htb]
\centering\includegraphics[width=0.75\linewidth]{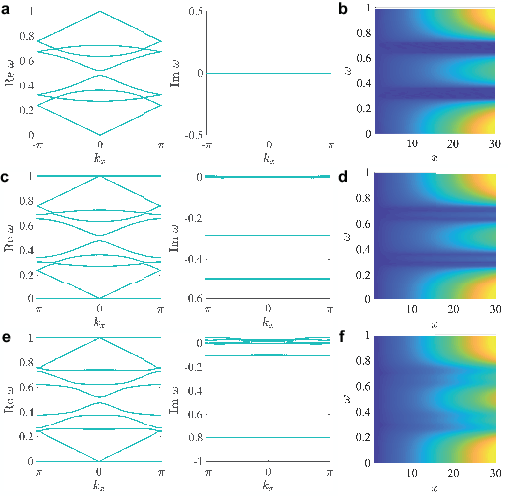}
\caption{\label{fig_simulation}{\bf Photonic band structure (a,c,e) and transmission spectrum (b,d,f) calculated from the transfer-matrix method}. With the given wave vector $k_x\in[-\pi,\pi]$, the real and imaginary parts of the normalized frequency $\omega$ can be solved from the eigenvalue equation in Eq. (\ref{eigenQ}). (a,b) Hermitian case with parameter $\kappa=0$. (c,d) Non-Hermitian case with parameter $\kappa=0.3$. (e,f) Non-Hermitian case with parameter $\kappa=0.7$.}
\end{figure}

Figure \ref{fig_simulation} demonstrate the numerical results of band structure and transmission spectrum calculated from transfer-matrix method. Figures \ref{fig_simulation}(a,b) demonstrates the results under the condition of $\kappa=0$, corresponding to the Hermitian scenario. In this case, under the given $k_x\in[-\pi,\pi]$, the photonic bands structure only possesses real parts with no imaginary components. Observing the variation of the electric field amplitude with the cell position $x$ at different frequencies $\omega$, it is evident that photonic band gaps exist around frequency $\omega = 0.32$ and $\omega = 0.68$. Figure \ref{fig_simulation}(c,d) and (e,f) represents the non-Hermitian scenario with $\kappa=0.3$ and $\kappa=0.7$, where gain and loss are introduced. We found that the photonic bands exhibit a phenomenon similar to band inversion in topological insulators nearby $\omega = 0.32$ and $\omega = 0.68$. The electric field amplitude reveals that new bands emerge at the original photonic band gaps, indicating the presence of topological boundary modes. In summary, the introduction of non-Hermitian terms endows our coupled resonator optical waveguides with the characteristics of topological phase transitions.

\section{Conclusion}
In summary, we reveals four-fold non-Hermitian phases at a mathematically level, where phase I exhibits only topological boundary modes, phase II displays both topological boundary modes and skin modes, phase III exhibits only skin modes, and phase IV cannot manifest any boundary modes. By deriving non-Hermitian winding numbers, the existence or non-existence condition of topological boundary modes are analytically expressed, consistent with the numerical results obtained through the iterative Green’s function method. Combining with the study on non-Hermitian skin effects, we rigorously establish the four-fold phase diagram. We also design an array of coupled resonator optical waveguides. The introduction of non-Hermiticity in our photonic structure induces a phenomenon similar to band inversion in topological insulators, indicating the presence of topological boundary modes in the photonic bands.

\appendix    
\setcounter{equation}{0}
\renewcommand\theequation{A.\arabic{equation}} 

\section{The iterative Green's function method}
\label{sec_green}
This section demonstrates the detailed calculation of the Green's function through iterative approach. The iterative Green's function method starts with the Hamiltonian with nearest inter-cell interaction, described as
\begin{equation}
\hat{H} = \sum_{n}[\bm{c}^{\dagger}_{n}M\bm{c}_{n}+\bm{c}^{\dagger}_{n}J_L\bm{c}_{n+1}+\bm{c}^{\dagger}_{n+1}J_R^{\dagger}\bm{c}_{n} ],
\end{equation}
where $M$ is the on-site matrix, and $J_L$, $J_R^{\dagger}$ are the inter-cell hopping matrices. Therefore, our non-Hermitian model should be regarded as a lattice with $N/2$ supercells, and each supercell contains four sites. The on-site matrix of supercell is given by
\begin{equation}
\begin{aligned}
M  =
\left(\begin{matrix}
\varepsilon_0 	&t_0			&0				&0\\
t_0				&\varepsilon_0	&t_{1L}				&0\\
0 				&t_{1R}			&\varepsilon_0	&t_0\\
0 				&0				&t_0			&\varepsilon_0
\end{matrix}\right),
\end{aligned}
\end{equation}
and the hopping matrices between the supercells are given by
\begin{equation}
\begin{aligned}
J_L  =
\left(\begin{matrix}
0 				&0				&0				&0\\
t_2				&0				&0				&0\\
0 				&0  			&0				&0\\
t_{1L}			&0				&t_2			&0
\end{matrix}\right),\ 
J_R^{\dagger}  =
\left(\begin{matrix}
0 				&t_2			&0				&t_{1R}\\
0				&0				&0				&0\\
0 				&0  			&0				&t_2\\
0				&0				&0   			&0
\end{matrix}\right).
\end{aligned}
\end{equation}
The Green's function $G(\omega)$ is defined as
\begin{equation}
\begin{aligned}
(\omega-H)G(\omega)=\mathbb{I},
\end{aligned}
\end{equation}
where $\omega = E+\ii \eta$ and $\eta\to 0^{+}$. 
 
The iterative approach is used to reduced the computational expense on $G(\omega)$. The method involves replacing each supercell plus its two neighbors by an effective layer, and this replacement can be repeated until the residual interaction between the effective layers can be ignored. Detailed proof of iterative Green's function method is shown in Ref.[\citenum{Green2018}], and the algorithm procedure is illustrated as below. $\varepsilon_b$ and $\varepsilon_{s,s'}$ are the effective intra-layer matrix of the bulk and the two boundaries, which are initialized as the on-site matrix $M$. $\alpha$ and $\beta$ are the residual interaction between layers, initialized as the hopping matrix $J_L$ and $J_R^{\dagger}$. $\eta$ is an infinitely small quantity to avoid singular matrix, and $error$ determines the convergence criteria. Here, we set $\eta =1\times10^{-3} $ and $error =1\times10^{-10}$. Finally, $A_b$ and $A_{s,s'}$ are the calculated bulk spectral function and surface spectral function respectively, as shown in Fig.\ref{fig3_green} in main text.

\begin{algorithm}[h]
    \caption{Iterative Green's Function Method }
    \LinesNumbered
    Initialization: $\varepsilon_b^{(0)} =\varepsilon_s^{(0)}=\varepsilon_{s'}^{(0)}=M,\ \alpha^{(0)} = J_L,\ \beta^{(0)} = \jr$   \\
    \For{$E$ = $\min\{E\}$ to $\max\{E\}$}{   
\For{$x$ = 0 to itermax-1 }{  
	$\varepsilon_b^{(x+1)} = \varepsilon_b^{(x)}+\alpha^{(x)}(E+\ii\eta-\varepsilon_b^{(x)})^{-1}\beta^{(x)}+\beta^{(x)}(E+\ii\eta-\varepsilon_b^{(x)})^{-1}\alpha^{(x)}$\\
	$\varepsilon_s^{(x+1)} = \varepsilon_s^{(x)}+\alpha^{(x)}(E+\ii\eta-\varepsilon_b^{(x)})^{-1}\beta^{(x)}$\\
$\varepsilon_{s'}^{(x+1)} = \varepsilon_{s'}^{(x)}+\beta^{(x)}(E+\ii\eta-\varepsilon_b^{(x)})^{-1}\alpha^{(x)}$\\
$\alpha^{(x+1)} = \alpha^{(x)}(E+\ii\eta-\varepsilon_b^{(x)})^{-1}\alpha^{(x)}$\\
$\beta^{(x+1)}= \beta^{(x)}(E+\ii\eta-\varepsilon_b^{(x)})^{-1}\beta^{(x)}$\\
        \eIf{$\Vert \alpha^{(x+1)} \Vert<$ error} {
            break
        } {         
        }
    }
$A_b(E)=-\frac{1}{\pi}{\rm Im\ [Tr}\ (E+\ii\eta-\varepsilon_b)^{-1}]$\\
$A_s(E)=-\frac{1}{\pi}{\rm Im\ [Tr}\ (E+\ii\eta-\varepsilon_s)^{-1}]$\\
$A_{s'}(E)=-\frac{1}{\pi}{\rm Im\ [Tr}\ (E+\ii\eta-\varepsilon_{s'})^{-1}]$\\
    }
\end{algorithm}

\acknowledgments 
 
The authors thank for the support from the National Natural Science Foundation of China (Grant Nos. 60578047, 61427815), and the Natural Science Foundation of Shanghai (Grant Nos. 17ZR1402200, 13ZR1402600).

\providecommand{\noopsort}[1]{}\providecommand{\singleletter}[1]{#1}%

\bibliographystyle{spiebib} 

\end{document}